\documentclass[conference,onecolumn,draftcls]{IEEEtran}
\usepackage{graphicx}
\usepackage{amsthm,amsmath}
\theoremstyle{definition} 

\newcommand{\bold}{\,\textbf}
\newcommand{\eq}[1]{\,\begin{equation}
                   #1 
                   \end{equation}
}

\newcommand{\morabba}[1]{\,\begin{flushright}
 \Rectsteel \\
\end{flushright}}

\newcommand{\eqq}[2]{\,\begin{equation} \label{#2}
                   #1 
                   \end{equation}
}

\title{Growing a Network on a Given Substrate}

\author{\IEEEauthorblockN{Babak Fotouhi and Michael G. Rabbat}
\IEEEauthorblockA{Department of Electrical and Computer Engineering\\
McGill University, Montr\'eal, Qu\'ebec, Canada\\
Email: babak.fotouhi@mail.mcgill.ca, michael.rabbat@mcgill.ca}
}

\begin{document}

\maketitle

\begin{abstract}
 Conventional studies of network growth models mainly look at the steady state degree distribution of the graph. 
Often long time behavior is considered, hence the initial condition is ignored. In this contribution, the time evolution of the degree distribution is the center of attention. We consider two specific growth models; incoming nodes with uniform and preferential attachment, and the degree distribution of the graph for arbitrary initial condition is obtained as a function of time. This allows us to characterize the transient behavior of the degree distribution, as well as to quantify the rate of convergence to the steady-state limit.
\end{abstract}

\section{Introduction}

The network paradigm is ubiquitous, including many fields such as epidemiology \cite{epid}, sociology \cite{ gran, Wasser, degenne}, 
economics \cite{econ1, econ2}, management \cite{harary} and marketing \cite{market1, martins, delre}, to name a few. 
The dynamics of the phenomena under consideration usually depends on the topological properties of the network, drawing attention to its formation process.
Network formation models for static population has been studied mostly under the category of random graphs with Poisson distribution \cite{Ballabos}.

There are ample applications for which the network evolves through time. In fact most real networks do, trivial examples being the society, the world-wide web, scientific citation 
networks, etc. Examples of conventional models for growing graphs are Small-World networks \cite{Watts1} and the so-called preferential attachment models
\cite{Barabasi1,Drog1}, usually associated with scale-free properties. 
In those models the long-time behavior of the graph (the thermodynamic limit) is of interest and the asymptotic distribution 
is investigated for which the initial conditions can be neglected \cite{Newman1,Red1,Red2}. 

On the contrary, one might be interested in the short-time regime of the growth process. One encounters this in population growth
 (where incoming nodes are born or are new immigrants that join the pre-existing network of a country) or
any process explicable in terms of the growth mechanism of the structure, 
 and wherever extrapolation of the time evolution of the graph might be of interest. On the other hand, although the eventual 
behavior is known, one might ask how fast does the network reach the steady-state, to  make sense of the equilibrium assumptions accordingly.


In this paper we derive closed form expressions for the expected degree distribution over time as a network evolves. We focus on two particular growth models. Both models begin with an initial network (the \emph{substrate}), and nodes are added to the network as time advances. In the first model, termed uniform selection, new nodes connect to one or more existing nodes without preference (i.e., drawn uniformly from the pool of existing nodes). In the second model, new nodes connect to one existing node by following the preferential attachment model~\cite{Barabasi2}, whereby the neighbor of the new node is drawn with probability proportional to its present number of neighbors. For both models we derive closed form expressions for the expected degree distribution as a function of time. These allow for characterization of the rate of convergence to the limiting distributions. In both models, the limiting distributions do not depend on characteristics of the initial substrate. 

The rest of this extended abstract is organized as follows. Section~\ref{sec:uniform} presents results for the uniform attachment model, and Section~\ref{sec:preferential} presents results for the preferential attachment model. Both 
of these sections are complemented with numerical simulations verifying the theoretical expressions we derive. Section~\ref{sec:conclusion} offers concluding remarks.

\section{Uniform Linking} \label{sec:uniform}

In this section we assume the new nodes pick single or multiple existing nodes and connect to them. 
They choose the destination nodes uniformly at random, i.e., regardless of their degree. It is assumed that the existing nodes at each timestep do not delete or add links. 

\subsection{Single Connection}

The model begins with an initial network, which we refer to as the substrate, at time $t=0$. We assume that the substrate contains $N(0)$ nodes, and the number of nodes in the substrate having degree $k \ge 1$ is given by $N_k(0) = n_k$. We also assume that the substrate is connected. If this is not the case, then under the uniform single connection model, the limiting graph will have the same number of components as the substrate, and new nodes join one component with a probability proportional to the current number of nodes in that component.

Suppose new nodes are born at a rate $\alpha$ and connect to an existing node which is drawn uniformly at random from the existing population, so at each time step $\alpha$ new nodes 
with degree 1 are added to the network. SInce there are $N(0)$ nodes at the outset, the total number of nodes in the graph after time $t$ is $N(0)+\alpha t$. Let $N_k(t)$ denote the expected number of nodes having degree $k$ at time $t$. We seek the time evolution of the quantities $\{N_k(t)\}$ in order to find an expression for the degree distribution by dividing by the total population at time $t$ afterwards. 
To do so, it is expedient to use the rates of transition between states to study the dynamics \cite{stat1}. According to the uniform attachment dynamics, the expected degree distribution evolves as follows for $t > 0$,
\eq{
\begin{cases}
N_k (t+\Delta t)=N_k(t) + \frac{\alpha }{N(0)+\alpha t} (N_{k-1}-N_k)  & k\neq 1 \\
N_1(t+\Delta t)=N_1(t) - \frac{\alpha }{N(0)+\alpha t} + \alpha \Delta t & k=1.
\end{cases}
}
These dynamics can be compactly represented by the following differential equation,
\eq{
\dot{N_k}=\frac{\alpha }{N(0)+\alpha t} (N_{k-1}-N_k)+\alpha \delta_{k,1}, \label{eqn:diffeq}
}
where $\dot{N_k}$ is the first derivative of $N_k(t)$ with respect to time, $\delta_{k,1}$ is the Kronecker delta function (i.e., $\delta_{k,1} = 1$ if $k=1$, and $\delta_{k,1} = 1$ otherwise), and dependence on time is suppressed to simplify the presentation. Since the initial network is connected by assumption and every new node connects to one existing node, we have that $N_0(t) = 0$ for all $t \ge 0$.

In order to solve~\eqref{eqn:diffeq}, we will use the generating function $\psi(z,t)=\sum_k z^{-k} N_k(t)$. We get
\eq{
\frac{\partial \psi(z,t)}{\partial t}=\frac{\alpha }{N(0)+\alpha t}(z^{-1}-1)\psi(z,t)+\frac{\alpha}{z},
}
which is solved by multiplying both sides by the integration factor $(t+\frac{N(0)}{\alpha})^{\frac{1}{z}-1}$. The result is
\eq{
\psi(z,t)=\frac{N(0)+\alpha t}{z(2-\frac{1}{z})}+C_z(t+\frac{N(0)}{\alpha})^{\frac{1}{z}-1},
}
and incorporating the initial condition gives
\eq{
\psi(z,t)=\frac{N(0)+\alpha t}{z(2-\frac{1}{z})}+\frac{N(0)}{N(0)+\alpha t} \left[\psi(z,0) (1+\frac{\alpha t}{N(0)})^{\frac{1}{z}}\right]-
\frac{N(0)^2}{N(0)+\alpha t} \frac{1}{z(2-\frac{1}{z})}\left(1+\frac{\alpha t}{N(0)}\right)^{\frac{1}{z}}.
}

Recalling that in the initial (substrate), $N_k(0) = n_k$ nodes have degree $k$, we find that the solution of $N_k(t)$ is given by,
\eq{
N_k(t)=\frac{1}{2^k} (N(0)+\alpha t) + \frac{N(0)^2}{N(0)+\alpha t} \left\{n_k * \frac{\left[ \ln \left( 1+\frac{\alpha t}{N(0)} \right) \right]^k}{k!} \right\}
-\frac{N(0)^2}{N(0)+\alpha t} \left\{ \left( \frac{1}{2^k} \right) * \frac{\left[ \ln \left( 1+\frac{\alpha t}{N(0)} \right) \right]^k}{k!} \right\},
}
where $*$ denotes the convolution operator. Dividing by the total number of nodes at time $t$ gives the expected degree distribution,
\eqq{
p_k(t)=\frac{1}{2^k}  + \left(\frac{N(0)}{N(0)+\alpha t}\right)^2 \left\{n_k\star \frac{\left[ \ln \left( 1+\frac{\alpha t}{N(0)} \right) \right]^k}{k!} \right\}
-\left(\frac{N(0)}{N(0)+\alpha t}\right)^2 \left\{ \left( \frac{1}{2^k} \right) \star \frac{\left[ \ln \left( 1+\frac{\alpha t}{N(0)} \right) \right]^k}{k!} \right\}.
}{result1}

Note that as $t \rightarrow \infty$, the term involving initial conditions vanishes, and the asymptotic behavior of the degree distribution is as follows
\eq{
\lim_{t \rightarrow \infty} p_k(t) =\frac{1}{2^k}.
}
This matches the asymptotic behavior previously found in \cite{Bollobas2,Red2}. 

\begin{figure}[ht]
  \centering
  \includegraphics[width=4in]{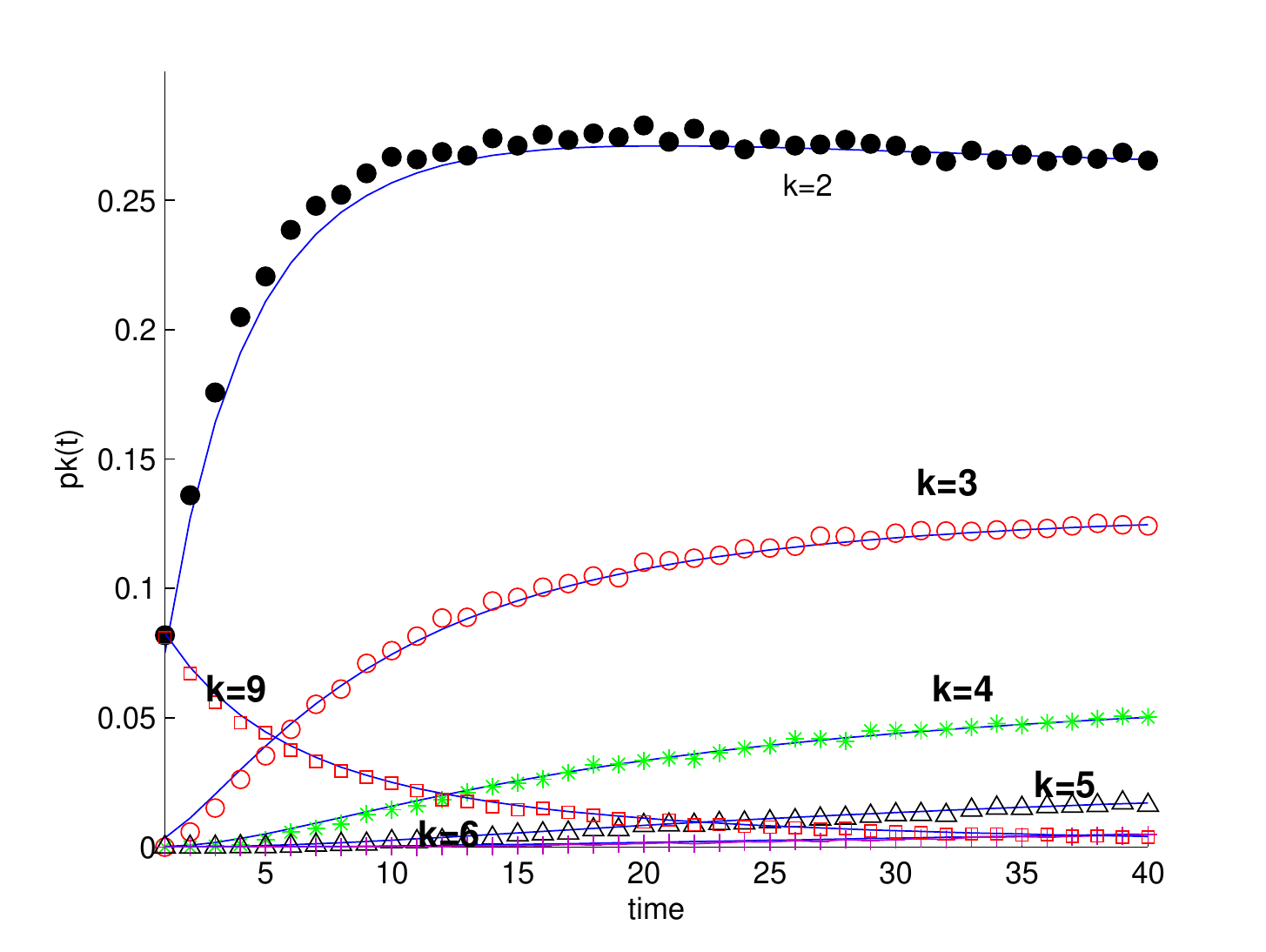}
  \caption[Figure 1]%
  {$p_k(t)$ for different values of $k$ as a function of time for the star topology in uniform linking scheme. Solid lines represent the theoretical prediction. }
\label{fig1}
\end{figure}

\begin{figure}[ht]
  \centering
  \includegraphics[width=4in]{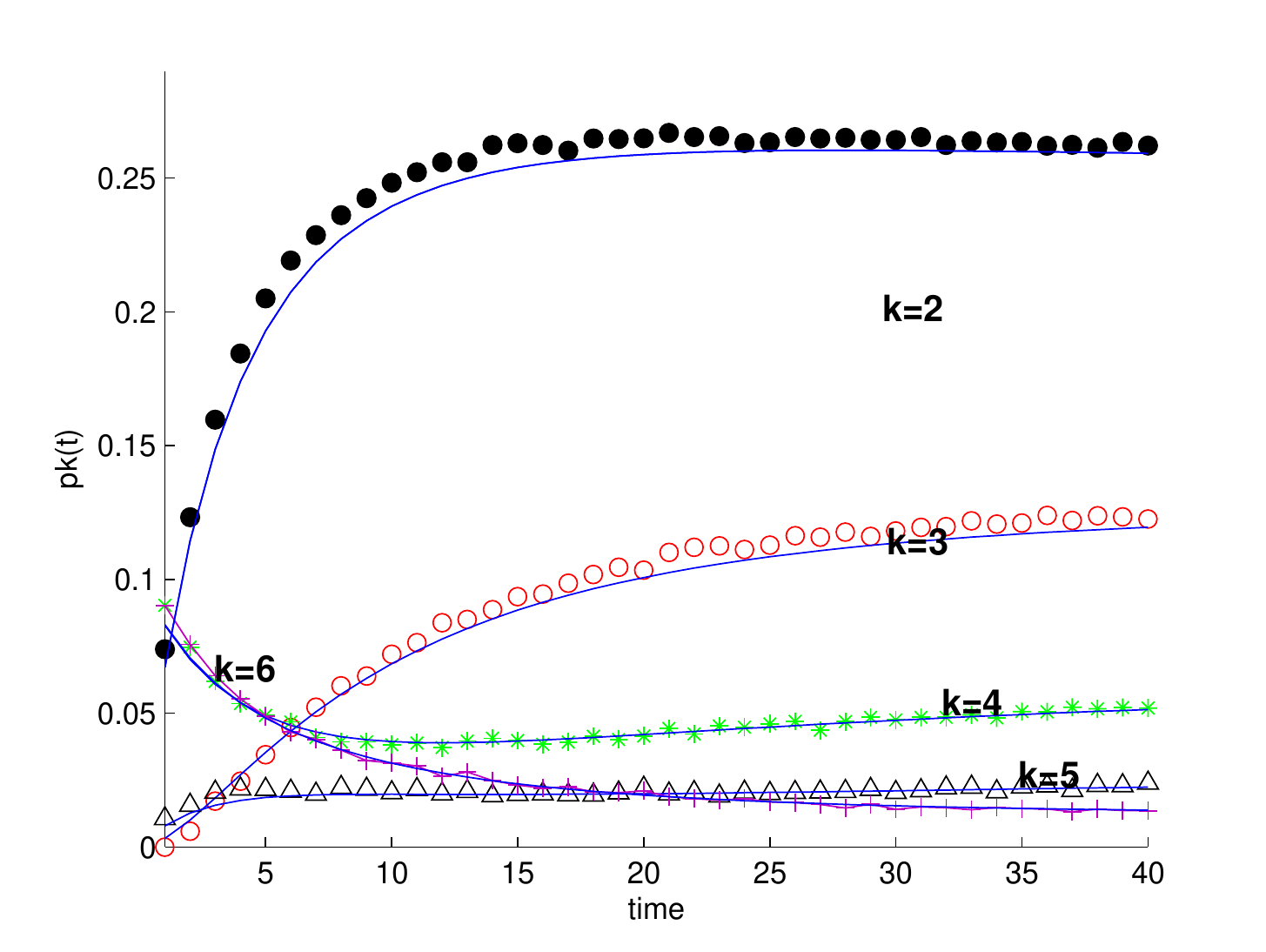}
  \caption[Figure 2]%
  {$p_k(t)$ for different values of $k$ as a function of time for the double-star topology in uniform linking scheme. Solid lines represent the theoretical prediction.}
\label{fig2}
\end{figure}

As an illustrative example, let us take the substrate to be a start with $N(0)$ nodes; i.e., there is one central node with degree $N(0)-1$, and all other nodes have a single edge connecting to the central node. We simulate this scenario with $N(0)=10$. Figure~\ref{fig1} shows the fraction of nodes with degrees $k=2$ -- $6$ and $9$. The solid lines show our theoretical expression, and the scatter points for each curve are the average over $10000$ Monte Carlo simulations. We use $\alpha=1$. Figure~\ref{fig2} shows similar results but for using a double-star for the initial substrate. The double star has two central nodes which each connect to each other, and each central node has a number of satellites. In our simulation, $N(0)=10$ and the central nodes have degrees $4$ and $6$. As predicted, the limiting degree distribution is independent of the distribution of the initial substrate.

\subsection{Multiple Connections}

Now, suppose that each newly-born node connects to $\beta$ existing nodes drawn uniformly at random from the existing population. Note that this model potentially yields very different results. In the former case (single connection), if the substrate begins as a disconnected graph then, with probability 1, the limiting graph contains the same number of connected components as the substrate. On the other hand, in the uniform multiple destination model there is non-zero probability that each new node will form a bridge between two components.

Under the uniform multiple connection model, the differential equation for $N_k(t)$ now becomes
\eq{
\dot{N_k}=\frac{\beta \alpha }{N(0)+\alpha t} (N_{k-1}-N_k)+\alpha \delta_{k,\beta} ~.
}
The generating function is of the form
\eq{
\psi(z,t)=\psi(z,0) \left( \frac{N(0)}{N(0)+\alpha t} \right) ^{\beta(1-\frac{1}{z})}
- \frac{N(0)}{z^{\beta}[1+\beta(1-\frac{1}{z})]} \left( \frac{N(0)}{N(0)+\alpha t} \right) ^{\beta(1-\frac{1}{z})}
+ \frac{N(0)+\alpha t}{z^{\beta}[1+\beta(1-\frac{1}{z})]}, 
}
and so the populations are given by
\begin{multline}
N_k(t)=
\frac{N(0)+\alpha t}{\beta} \left( \frac{\beta}{\beta+1} \right)^{k-\beta+1} u(k-\beta)
+ \left[ \frac{N(0)^{\beta+1}}{(N(0)+\alpha t)^{\beta}} \right]    \left\{ n_k * \frac{\left[\beta  \ln \left( 1+\frac{\alpha t}{N(0)} \right) \right]^k}{k!} \right \}\\
-\left[ \frac{N(0)^{\beta+1}}{\beta(N(0)+\alpha t)^{\beta}} \right]
\left\{ \left( \frac{\beta}{\beta+1} \right)^{k-\beta+1} u(k-\beta)  * \frac{\left[\beta  \ln \left( 1+\frac{\alpha t}{N(0)} \right) \right]^k}{k!} \right\},
\end{multline}
were $u(x)$ is the Heaviside step function (i.e., $u(x) = 0$ for $x < 0$, and $u(x) = 1$ for $x\ge0$). Normalizing by the total population at time $t$, we arrive at the degree distribution:
\begin{multline}
p_k(t)=
\frac{1}{\beta} \left( \frac{\beta}{\beta+1} \right)^{k-\beta+1} u(k-\beta)
+ \left( \frac{N(0)}{N(0)+\alpha t} \right)  ^{\beta+1}     \left\{ n_k * \frac{\left[\beta  \ln \left( 1+\frac{\alpha t}{N(0)} \right) \right]^k}{k!} \right \}\\
-\left( \frac{N(0)}{N(0)+\alpha t} \right)  ^{\beta+1}  \frac{1}{\beta}
\left\{ \left( \frac{\beta}{\beta+1} \right)^{k-\beta+1} u(k-\beta)  * \frac{\left[\beta  \ln \left( 1+\frac{\alpha t}{N(0)} \right) \right]^k}{k!} \right\}.
\end{multline}
Asymptotically, the second and third terms vanish, and we have
\begin{equation}
p_k(t) \sim \frac{1}{\beta} \left( \frac{\beta}{\beta+1} \right)^{k-\beta+1} u(k-\beta).
\end{equation}
Of course, since every new node connects $\beta$ existing nodes, as $t\rightarrow \infty$, all nodes have degree at least $\beta$. Likewise, taking $\beta = 1$, we recover the same asymptotic distribution as in the previous section.

\begin{figure}[ht]
  \centering
  \includegraphics[width=4in]{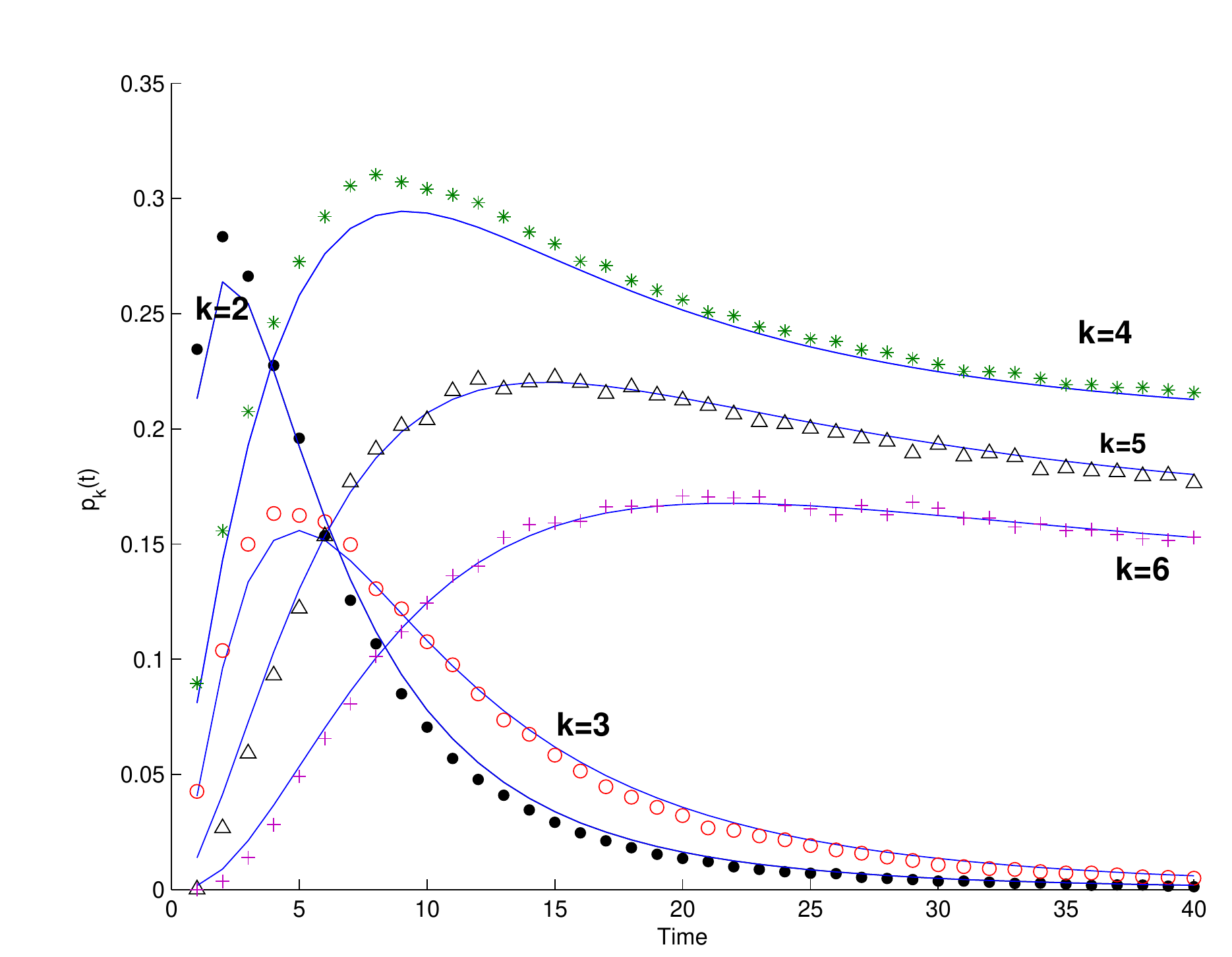}
  \caption{$p_k(t)$ for different values of $k$ as a function of time for the star topology in uniform linking scheme where each new node connects to $\beta=4$ existing nodes. Solid lines represent the theoretical prediction. }
\label{star_beta4}
\end{figure}

\begin{figure}[ht]
  \centering
  \includegraphics[width=4in]{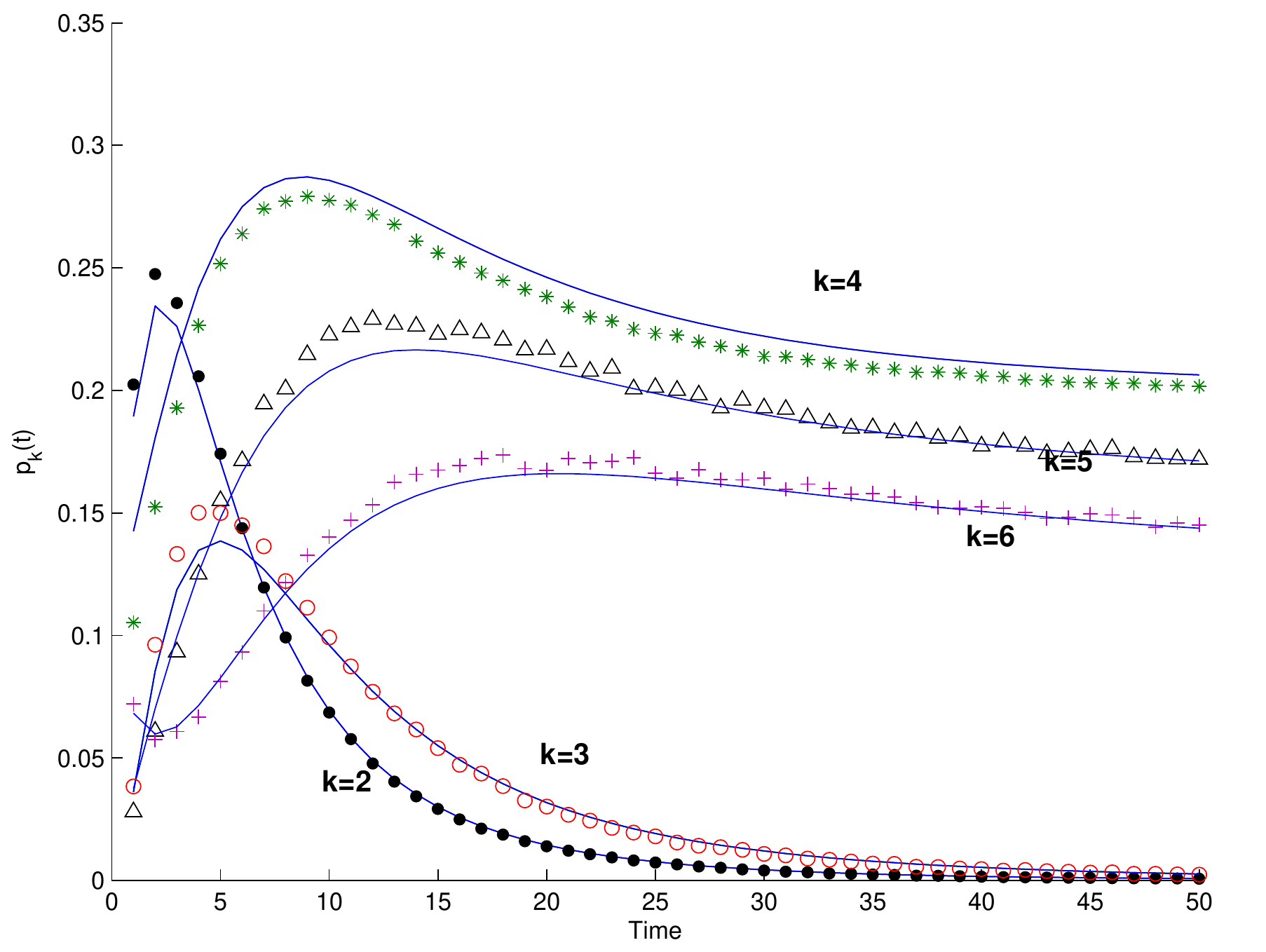}
  \caption{$p_k(t)$ for different values of $k$ as a function of time for the double-star topology in uniform linking scheme where each new node connects to $\beta=4$ existing nodes. Solid lines represent the theoretical prediction.}
\label{dstar_beta4}
\end{figure}

Simulation results are depicted Figures~\ref{star_beta4} and \ref{dstar_beta4} for the same two substrates as used above, and with $\beta=4$.

\section{Preferential Linking} \label{sec:preferential}

It might be the case that nodes are not impartial to the status of the nodes they want to select for attachment. There are situations exhibiting 
the Mathew effect, where popularity is self-reinforcing (rich-gets-richer). Scientific citation networks are examples of that \cite{price,leij,newman2}. 
Suppose each new node connects to an existing node drawn with a probability proportional to a function of its current degree. Any increasing function of degree can be used to model preferential attachment; here we focus on the case where probabilities are linearly proportional to degree~\cite{Barabasi2}. 
The evolution of $N_k(t)$ is given by the differential equation
\eqq{
\dot{N_k}=\frac{\alpha }{\sum_k k N_k} ((k-1)N_{k-1}-k N_k)+\alpha \delta_{k,1}.
}{Nkdot2}
Recall that $N_k(t) = (N(0) + \alpha t) p_k(t)$. We use the generating function for $p_k(t)$ defined as
\eq{
\psi(z,t)=\sum_k p_k(t) z^{-k}.
}
Multiplying~\eqref{Nkdot2} by $z^{-k}$ and summing over $k$, we get
\eq{
(N(0) + \alpha t) \frac{\partial \psi}{\partial t} - \frac{\alpha (z-1)(N(0)+\alpha t)}{(N(0) \bar{k}_0+2 \alpha t)} \frac{\partial \psi}{\partial z}
= \alpha z^{-1} - \alpha \psi, \label{eqn:diffeq2}
}
where $\bar{k}_0$ is the average degree of the substrate. Equation \eqref{eqn:diffeq2} is a linear partial differential equation which has characteristic curve $(z-1)^2 (N(0) \bar{k}_0 + 2 \alpha t)= C$. Using this, the solution has the form 
\eq{
\psi(z,t)=\frac{\alpha \Phi(C)}{N(0)+\alpha t}+\frac{\frac{1}{2}}{N(0)+\alpha t} \left[ -2 \sqrt{C(N(0) \bar{k}_0 + 2 \alpha t)}+
(N(0) \bar{k}_0 + 2 \alpha t)+2C \ln (\sqrt{C}+\sqrt{(N(0) \bar{k}_0 + 2 \alpha t)}) \right]
}
where $\Phi(C)$ is uniquely determined by the initial conditions $N(0)$ and $\bar{k}_0$.

Define the constant
\eq{
c\equiv 1-\sqrt{\frac{N(0) \bar{k}_0}{N(0) \bar{k}_0+2 \alpha t}}.
}
The generating function simplifies to
\eq{
\psi(z,t)=\frac{N(0) \psi(\frac{z-c}{1-c},0)}{N(0)+\alpha t} + 
\frac{N(0) \bar{k}_0 + 2 \alpha t}{  N(0)+\alpha t} \sum_{k=1}^{\infty} z^{-k} \left(\frac{c^{k+2}}{k+2}-2\frac{c^{k+1}}{k+1}+\frac{c^k}{k} \right),
}
and the degree distribution is thus
\begin{multline}
p_k(t)=
\frac{N(0) \bar{k}_0 + 2 \alpha t}{ N(0)+\alpha t} \left(\frac{c^{k+2}}{k+2}-2\frac{c^{k+1}}{k+1}+\frac{c^k}{k} \right) \\ +
\left(\frac{N(0)}{N(0) + \alpha t} \right) 
\left\{ \sum_{m=1}^{k} n_m \binom{k-1}{m-1}  \left[\frac{N(0) \bar{k}_0}{N(0)\bar{k}_0+2\alpha t} \right]^{\frac{k}{2}} 
\left[ 1-\sqrt{\frac{N(0) \bar{k}_0}{N(0) \bar{k}_0+2 \alpha t}}  \right]^{k-m}   \right\}   .
\end{multline}
Note that $c \rightarrow 1$ as $t \rightarrow \infty$, and the second term on the right hand side also vanishes as $t\rightarrow \infty$. Thus, the asymptotic behavior is
\eq{
p_k(t) \sim 2 \left(  \frac{1}{k+2} - \frac{2}{k+1} + \frac{1}{k} \right),
}
which simplifies to
\eq{
p_k\sim \frac{4}{k (k+1) (k+2)}.
}

This asymptotic result was previously found in \cite{Red1}, which is equivalent to the $k^{-3}$ power law for the so-called scale-free graphs \cite{Barabasi1}.

\begin{figure}[ht]
  \centering
  \includegraphics[width=4in]{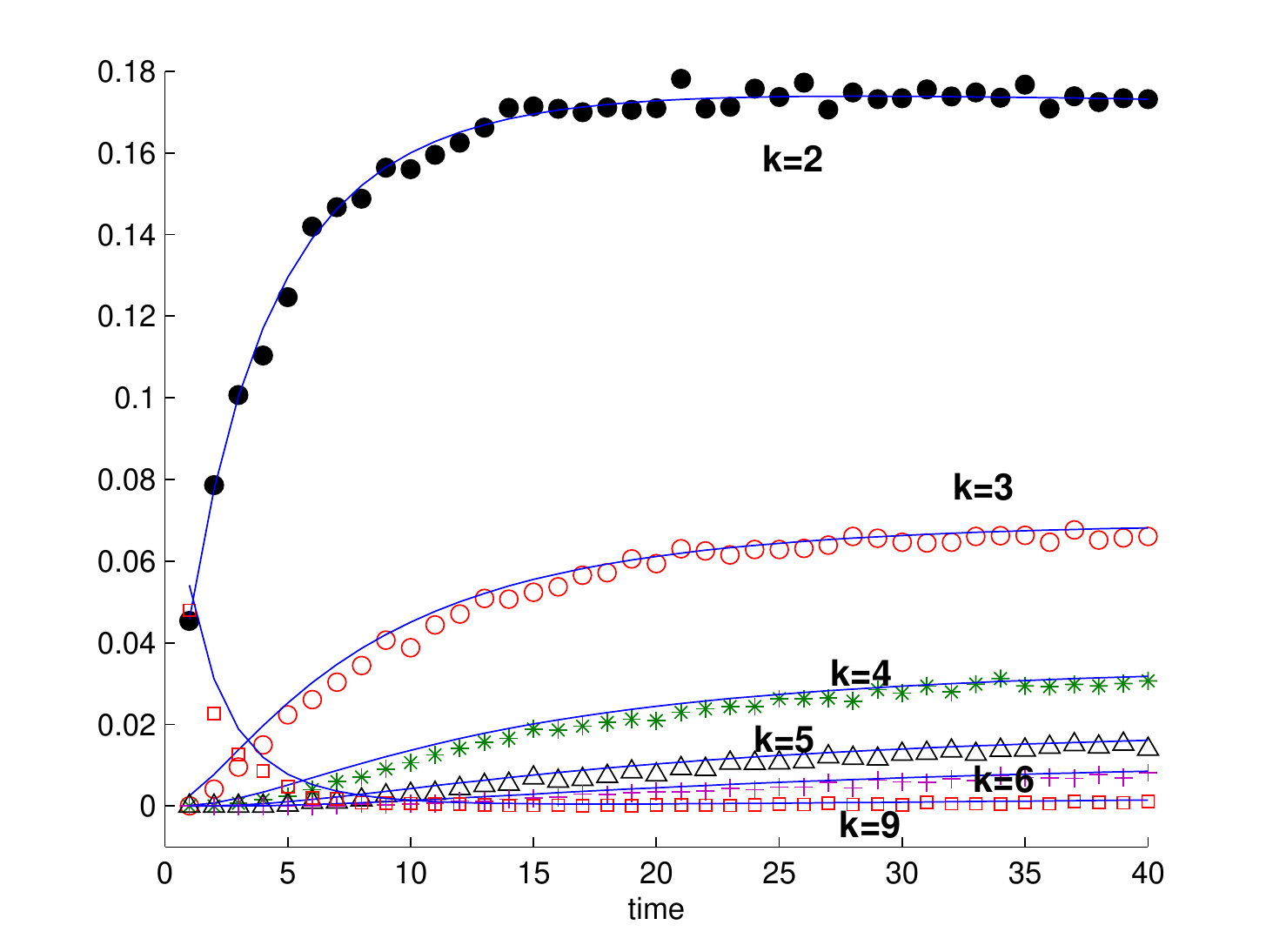}
  \caption[Figure 3]%
  {$p_k(t)$ for different values of $k$ as a function of time for the star topology in preferential linking scheme.}
\label{fig3}
\end{figure}

\begin{figure}[ht]
  \centering
  \includegraphics[width=4in]{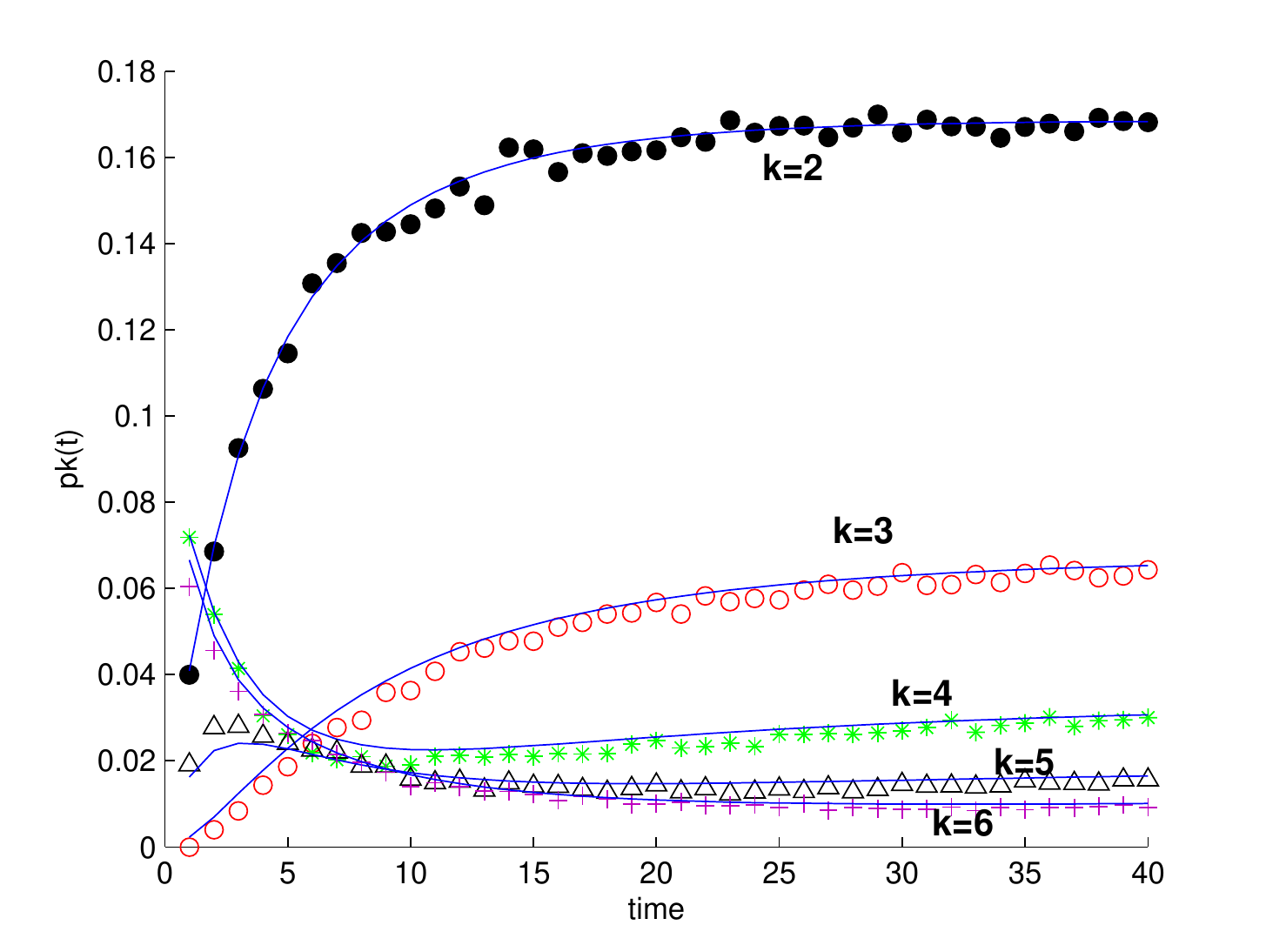}
  \caption[Figure 4]%
  {$p_k(t)$ for different values of $k$ as a function of time for the double-star topology in preferential linking scheme.}
\label{fig4}
\end{figure}

To verify this result we take the same initial graphs as before (start and double-star with $N(0)=10$). 
The analytical result is compared with the simulation results in Figures \ref{fig3} and \ref{fig4}. Again, the scatter points indicate averages over 10000 Monte Carlo trials with $\alpha = 1$.

\section{Summary} \label{sec:conclusion}

Previous studies of network growth models have primarily focused on the steady-state degree distribution. In this paper we find the exact degree distribution of a growing network as a function of time, 
for arbitrary initial conditions. This can be used to determine rates of convergence to the asymptotic limits. Understanding the short time behavior of graph evolution may also be useful in situations where the model parameters $\alpha$ and/or $\beta$ vary slowly with time. In these cases, the existing asymptotic analysis no longer applies, but it may be reasonable to treat the parameters as constant over short time scale and finding a piece-wise solution.

For the growth process, uniform and preferential linking of the newly born nodes are considered and theoretical predictions are compared with simulations. The full version of the paper will also contain results for a preferential attachment model where new nodes make connections to multiple existing nodes.

\end{document}